\documentclass[10pt,towcolumn,journal]{IEEEtran}
\usepackage{graphicx,amssymb,amsmath,amsthm,cite,cases}
\usepackage{adjustbox}

\usepackage{graphicx,amssymb,amsmath,amsthm,cite,cases}

\def\be{\begin{equation}}
\def\ee{\end{equation}}
\def\ben{\begin{eqnarray}}
\def\een{\end{eqnarray}}

\def\std{\mathrm{std}}

\def\PRD{\mathrm{PRD}}
\def\PRDB{\mathrm{PRD_B}}

\def\CR{\mathrm{CR}}

\def\CRD{\mathrm{{{CR}}{\,(1D)}}}
\def\CRDD{\mathrm{{{CR}}{\,(2D)}}}

\def\til{\tilde}

\def\vA{\mathbf{A}}
\def\vAr{\mathbf{A}^r}
\def\DCT{\widehat{\cal{C}}_{1c}}
\def\WT{\widehat{\cal{W}}_{1r}}
\def\vAt{\mathbf{B}}
\def\vAtr{\mathbf{B}^r}
\def\at{b}
\def\atr{b^r}
\def\vdel{\boldsymbol{{\delta}}}
\def\vat{\mathbf{b}}
\def\vatr{\mathbf{b}^r}
\def\vhb{\mathbf{h}}

\def\vs{\mathbf{s}}
\def\vst{\mathbf{\til{s}}}
\def\st{\til{s}}
\def\ellt{\til{\ell}}

\def\vc{\mathbf{c}}
\def\vct{\mathbf{\til{c}}}

\def\ct{\til{c}}
\def\vf{\mathbf{f}}
\def\vfr{\mathbf{f}^{\rm{r}}}

\def\vh{\mathbf{h}}

\def\vA{\mathbf{A}}

\title{Mixed-transform based codec for 2D compression of ECG signals}
 \author{Johan Chagnon and Laura Rebollo-Neira\\
Mathematics Department\\
Aston University\\
B3 7ET, Birmingham, UK}
\begin{document}
\maketitle
\begin{abstract}
A method for ECG compression, 
by imaging the record as a 2D array and 
implementing a transform lossy compression strategy,  
is advanced. The particularity of the proposed  
 transformation consists in applying a Discrete Wavelet 
Transform along one of the 
dimensions and the Discrete Cosine Transform along the 
other dimension. The performance of the  method is 
demonstrated on the MIT-BIH Arrhythmia database. 
Significant improvements upon the 1D version of  the
 codec, and on benchmarks for 2D ECG compression,  
are achieved.
\end{abstract}
\maketitle 
\section{Introduction}
Cardiovascular diseases (CVDs) are the number one 
cause of global death. 
The World Hearth Organization has estimated that 
17.9 million people died from CVDs
in 2016.  Over three quarters of these deaths take place  in low and middle income countries.  
A  major goal in the Sustainable 
Development Agenda of the United Nations is to reduce 
these figures one third by 2030. Within this agenda, 
prevention and routine controls play a central role.  
The electrocardiogram (ECG) is one of the most common 
tests in the diagnosis of CVDs. 
It goes without saying that techniques 
for safely compressing these types of data are essential 
to the development and support of clinical health care. 

In a recent publication \cite{LRN19} a 
method for effective high compression of ECG signals 
has been proposed. 
That method, which is applied on a 1D record, 
was shown to significantly improve  upon 
 benchmarks on the same database 
\cite{LKL11,MZD15,TZW18}. In this Communication we 
extend the compression technique in \cite{LRN19} to allow 
for its application 
on a 2D array constructed out of 1D ECG signal. 
It is demonstrated that, for the same level of 
distortion, the average compression performance 
on the MIT-BIH Arrhythmia database 
considerably improves in relation to the 1D processing. 
The method is also shown to 
 improve upon previous state of the art benchmarks 
concerning 2D ECG compression \cite{TSY05}. 
MATLAB software for reproducing results and facilitating 
future comparisons has been made 
available on a dedicated webpage \cite{webpage}.

\section{Proposed Coding Strategy}
A digital ECG signal represents a sequence of 
heartbeats, each of which is characterized by a 
combination of three graphical deflections, known 
as QRS complex, and the so called P and T waves. 
2D ECG compression relies on this 
feature. Our approach operates on raw data and consists of 
the following steps. 
\begin{itemize}
\item[A)] 1D to 2D conversion by segmentation  and 
alignment of  heartbeats.
\item[B)] Application of a Discrete Wavelet
Transform (DWT) along the direction of the segmented beats 
 and the Discrete Cosine Transform (DCT) along the 
perpendicular direction.
\item[C)] Quantization, organization, and entropy 
coding of the information needed to recover the ECG record
from the compressed file.
\end{itemize}
\subsection{1D to 2D conversion}
\begin{figure}[ht!]
\begin{center}
\includegraphics[width=5.5cm]{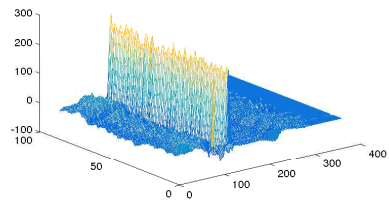}
\end{center}
\caption{2D array produced by segmentation and alignment of
 heartbeats.}
\label{FIG1}
\end{figure}

The conversion of the 1D ECG record into a 2D array 
requires segmentation and alignment of heartbeats. 
We implement this step as done in \cite{TSY05}.
Firstly a QRS detection algorithm is applied 
to locate the R peaks. Each of these peaks is 
aligned with the previous one.
 Since the length of the heartbeats are
not uniform, a regular array $\vA$
is obtained by padding rows with zeros. 
The duration of the heartbeats are stored as components of a
vector, say $\vh$, which has to be passed on to the 
decoder. Fig.\ref{FIG1} illustrates a 2D array of size $86 \times 359$ 
produced from a short 1D ECG record consisting of 25,000 samples. 

\subsection{Transformation of the 2D array}
\begin{figure}[ht]
\begin{center}
\includegraphics[width=6cm]{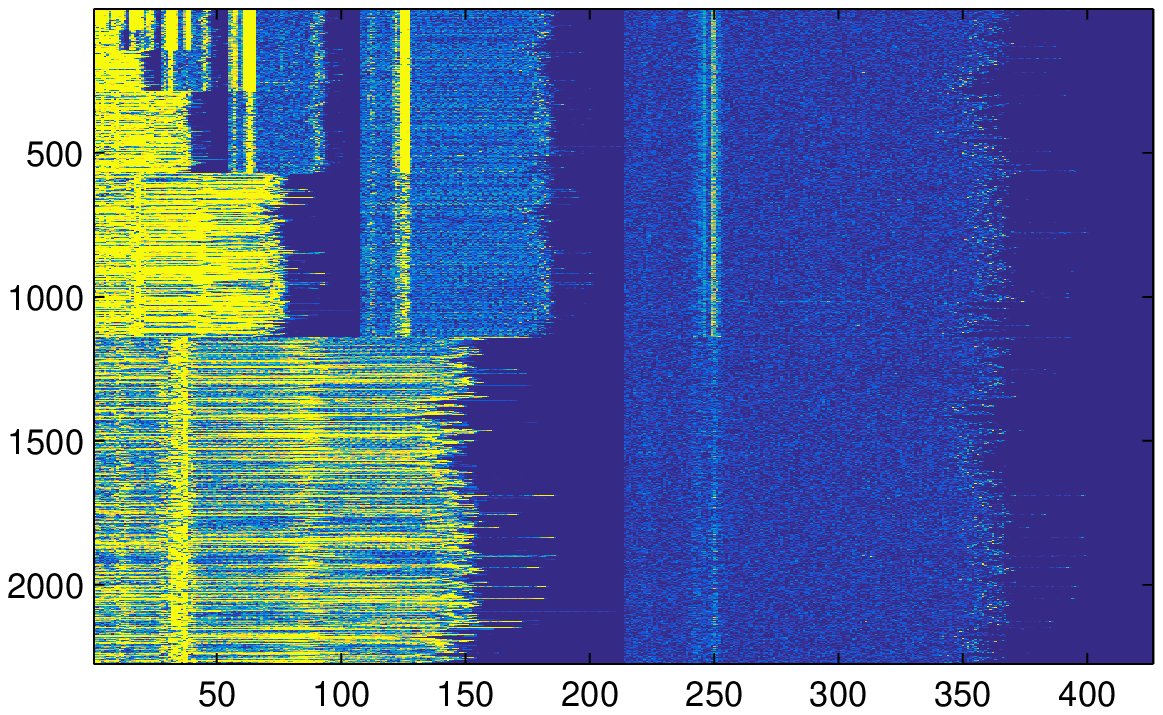}\\
\includegraphics[width=6cm]{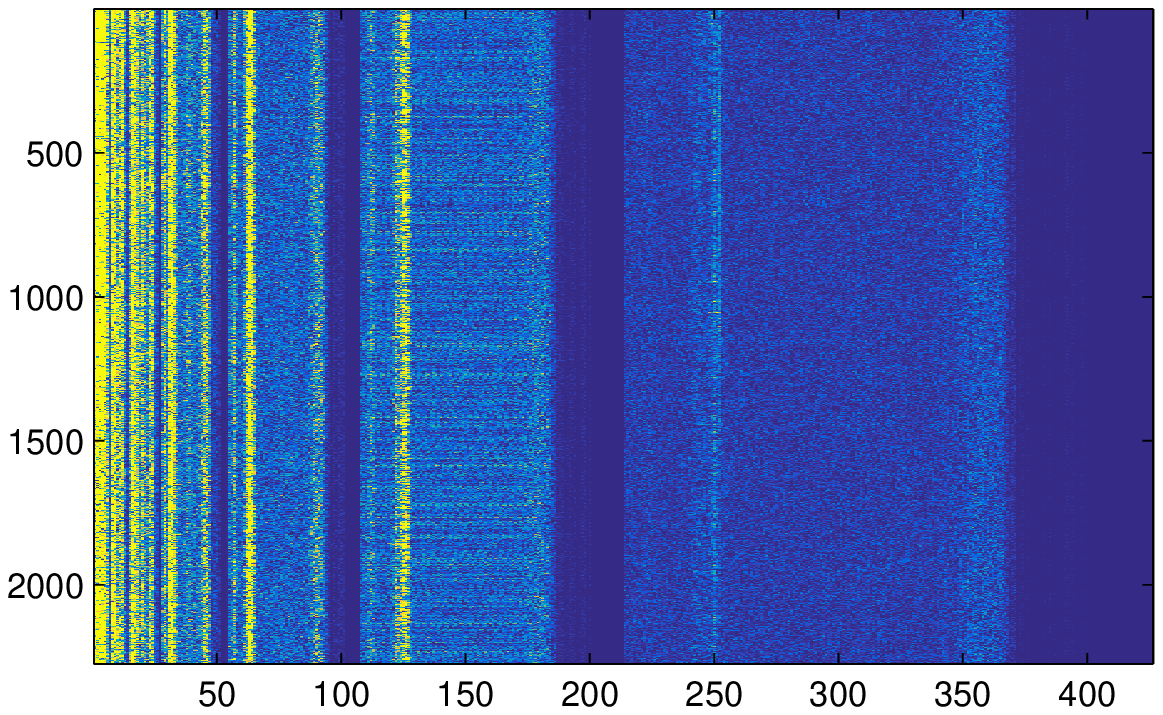}
\end{center}
\caption{Magnitude of the 2D  cdf97  DWT (upper graph)  and 
magnitude of mixed transform given in \eqref{At} (lower graph).}
\label{Fig2}
\end{figure}

This step introduces the distinctive feature of 
the proposed codec. Instead of applying a 2D wavelet 
transform on the array $\vA$,
 as it is done in other 2D ECG compression 
methods, e. g. \cite{TSY05}, we apply a different 
transform in each dimension. More precisely, we
apply the 1D cdf97 DWT   
 on the rows of $\vA$ and the 1D  DCT 
transform on the columns of $\vA$, i.e.,   
we create the transformed array $\vAt$ as follow:
\be
\label{At}
\vAt= \WT\DCT \vA, 
\ee
where $\WT$ indicates the 1D cdf97  DWT 
 operating on the rows of the array
 $\DCT \vA$ and $\DCT$ indicates the DCT transform 
operating on the columns of the array $\vA$. 
The convenience of performing this mixed transformation for 
encoding purposes becomes clear from the graphs 
of Fig.~\ref{Fig2}. The upper graph represents  
 the absolute value of the 
2D  cdf97  DWT of $\vA$ and the lower graph
the absolute value of the array $\vAt$ as given in 
\eqref{At}. The distribution of most significant 
elements in  $\vAt$ (lightest regions in the lower
graph) benefits the coding strategy described below.
\subsection{Encoding}
The encoding process begins by adopting a 
column-major order to express the $N \times M$ 
 array $\vAt$ as 
a 1D vector $\vat=(\at(1),\ldots,\at(N M))$. 
Hereafter the encoding proceeds 
as in  \cite{LRN19}. The components of $\vat$ 
 are converted to integer  numbers 
 by a mid-tread uniform quantizer as follows:
\be
\label{uniq}
\at^\Delta(i)= \lfloor \frac{\at(i)}{\Delta} +\frac{1}{2} \rfloor,\quad  i=1,\ldots,N M.
\ee
where $\lfloor x \rfloor$ indicates the largest
integer number
smaller or equal to $x$ and  $\Delta$ is the quantization
parameter. 
 
The absolute value of  
the elements  \eqref{uniq} are placed in 
a smaller vector, say $\vc= (c(1),\ldots,c(K))$, after the 
elimination of zeros. The signs are encoded 
separately in a vector $\vs=(s(1),\ldots,s(K))$
 using a binary alphabet 
(1 for + and 0 for -).

Assuming that 
the nonzero values in \eqref{uniq} occur at  
the positions $\ell_i,\ldots,\ell_K$, these 
indices are re-ordered in ascending
 order $\ell_{i} \rightarrow \til{\ell}_i,\,i=1,\ldots,K$,
which guarantees that
$\til{\ell}_i < \til{\ell}_{i+1},\,i=1,\ldots,K$.
This induces
a re-order in the coefficients,
$\vc \rightarrow \vct$ and
in the corresponding signs $\vs \rightarrow
\vst$.
Defining $\delta(i)=\ellt_i-\ellt_{i-1},\,i=2,\ldots,K$
the array
$\vdel=(\ellt_1, \delta(2), \ldots, \delta(K))$ stores
 the indices
$\ellt_1, \ldots, \ellt_K$ with unique recovery.

Finally the vectors $\vct, \vst, \vdel$, as well as  
the  length of the 
heartbeats $\vhb$, are compressed  using Huffman 
coding. The additional numbers which have to be 
passed to the decoder are: i) the quantization parameter 
$\Delta$ ii) the mean value of the 1D ECG record and 
iii) the size of the 2D array $\vA$. 
\section{1D ECG signal recovery}
At the decoding stage, after reverting  
 Huffman coding, the locations  $\ellt_1, \ldots, \ellt_K$ 
of the nonzero entries in the transformed array, 
 after quantization, are readily obtained. This allows the 
recovery of the array as follows. 
i) Set $\atr(i)=0,\, i=1,\ldots, N M$  and
$\atr(\ellt_i)= (2\st(i)-1)\ct(i)\Delta,\, i=1,\ldots, K$.  
ii) Reshape  the 
 vector $\vatr$ as a 2D array $\vAtr$ of size $N \times M$. 
The array $\vAr$ is recovered from $\vAtr$ inverting 
the $ \WT  $ and $ \DCT $ transformations (c.f.\eqref{At}).  
The re-conversion to the 1D signal say, $\vfr$, from the 
2D array $\vAr$ is straightforward using the 
heartbeat lengths in $\vhb$ and the mean value 
signal.

The achieved compression ratio $\CR$, which is defined as 
\be
\text{CR}=\frac{{\text{Size of the uncompressed file}}.}{{
\text{Size of the compressed file}}}
\ee
depends on the required quality of the 
recovered signal. This is assessed with respect to 
the standard $\PRD$ metric as given by 
\be
\PRD=\frac{\|\vf - \vfr\|}{\|\vf\|} \times 100 \%,\quad
\ee
where, $\vf$ is the original signal, 
 $\vfr$ is
the signal reconstructed from the compressed file and
$\| \cdot\|$ indicates the 2-norm. 
\section{Numerical Results}
For all the tests  the
 full MIT-BIH Arrhythmia database \cite{MITDB}, 
which contains 48 ECG records, is used. Each of these records consists of $N=650,000$ 11-bit samples at a frequency of 360 Hz.
At the QRS detection step a MATLAB implementation \cite{Sed14} of the Pan Tompkins algorithm was applied \cite{PT85}.

The comparison with 1D compression is realized with 
respect to the 1D version of the 
strategy adopted here, which 
has been shown in \cite{LRN19} 
to over-perform previously reported results \cite{LKL11,MZD15,TZW18}. 
Table~\ref{TABLE1} produces comparisons 
for $\PRD$ in the range [0.4 1]. The range 
[0.4 0.7]  corresponds 
to very low level distortion of the recovered signal. 
For $\PRD <0.4$ the proposed coding 
strategy is not effective. 
Nonetheless, it is worth noting that, because the approach is 
applied on raw data, requiring $\PRD <0.4$  
 implies to force the reproduction of the small high
 frequency noise present in all the records.
As discussed in \cite{LRN19},  
the decomposition of the cdf97 DWT in 4 levels produces the best 1D compression results on the MIT-BIH Arrhythmia database. 
In 2D, however, results improve by decomposing into 6 levels. Table \ref{TABLE1} shows the mean value $\CR$ and 
corresponding standard deviation (std) yielded by 
the 1D and 2D methods.
In both cases the quantization parameter has a different  
value for each record so as to achieve, 
for every record in the database, the 
sharp values of $\PRD$ given in 
the first row of Table~\ref{TABLE1}. 
Since the compressibility of the records 
is not uniform, fixing the same $\PRD$ for all the 
  records   
generates large dispersion in CR. This is 
 reflected in the std values.
%
\begin{table}[h]
\caption{Mean value Compression Ratios 
for  $\PRD$ values  sharply reproduced
 by every record in the database.}
\label{TABLE1}
\begin{center}
\begin{tabular}{||l||r|r|r|r|r|r|r||}
\hline \hline
$\PRD$& 1.00& 0.90& 0.80 & 0.70& 0.60& 0.50&0.40\\
\hline 
$\CRD$&42&39&35&32&28&24&19\\ 
$\std$&12&11&10& 9&8&7& 5 \\ 
$\CRDD$&85&73&61& 50& 39& 29& 20\\
$\std$&50&42&34& 28& 21& 15& 10\\
\hline \hline
\end{tabular}
\end{center}
\end{table}

Table \ref{TABLE3} produces comparison with  
results in Table VIII of \cite{TSY05}, for the whole 
database with reconstruction quality in our range 
 of interest.  The method in 
\cite{TSY05} employs a modified set partitioning 
in hierarchical trees (SPIHT) algorithm, which is shown 
to produce superior results than other image 
compressions techniques, including 
JPEG2000. 
The reported values of $\PRD$ are said to have been 
calculated after subtraction of a baseline of 1024 to 
the original data. The corresponding 
values are indicated here as $\PRDB$. The 
notation $\PRD$ is kept to indicate the $\PRD$ with respect to 
raw data.
\begin{table}[h]
\caption{Comparison with the results in TABLE VII of 
\cite{TSY05} for the whole database}
\label{TABLE3}
\begin{center}
\begin{tabular}{||c||c|c|c|c|c|c||}
\hline \hline
Method& $\PRDB$ & std& CR& std & $\PRD$ &std\\\hline 
\cite{TSY05}& 6.82 & -- & 30 & --  & --& --\\
1D &6.82&0.01&31 & 14 &0.77 &0.41\\
2D &6.82&0.01&58& 63&0.77&0.41\\ \hline 
\cite{TSY05} & 3.81 &-- & 20 &--& --&--\\
1D &3.81&0.01&19 & 10 &0.43 &0.23\\
2D &3.81&0.01&26&30&0.43&0.23\\\hline \hline
\end{tabular}
\end{center}
\end{table}
\section{Remarks and Conclusions}
\begin{figure}[ht!]
\begin{center}
\includegraphics[width=6cm]{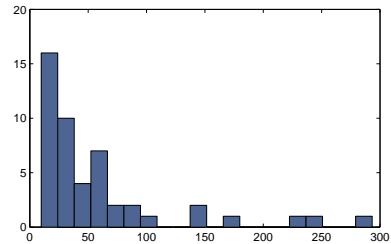}
\end{center}
\caption{Histogram of the CR obtained with the 2D approach 
 corresponding to $\PRDB=6.82$ for every record in the database.}
\label{FIG3}
\end{figure}

The numerical tests demonstrate that
the extension to operate in 2D  of the 1D 
compression strategy 
proposed in \cite{LRN19} yields,    
on the whole,  
very significant improvement in 
compression power. A distinctive feature of the 2D approach 
 is that, fixing quality, 
there is very large dispersion in the values of 
CR. This is due to the fact that compressing in 2D is 
 greatly beneficial for records of very regular morphology. 
Only some of the records in the 
 MIT-BIH Arrhythmia database possess this trait. 
 Fig.~\ref{FIG3} shows the histogram of CR for the 
2D approach and $\PRDB=6.82$. 

We are aware that the calculation of $\PRD$ subtracting a 
baseline of 1024 adopted in \cite{TSY05} has generated confusion leading to propagation of 
unfair comparison with values of $\PRD$ without subtraction 
of baseline. Table \ref{TABLE3} gives  
the two metrics. The quantization parameter $\Delta$, 
which controls quality, has been set differently 
for each record in 
oder to reproduce the sharp values of $\PRDB$ as those 
reported in \cite{TSY05}. The values of 
std are not reported in that publication.

The benefit of 2D compression comes at the expense of some 
 additional computation. With respect to the 1D implementation
 there is an extra time for QRS detection and for 
converting the 1D array into a 2D one. Furthermore, the 
Huffman coding step becomes more relevant in 
2D compression than it is in 1D.  However, as shown in 
Tables \ref{TABLE1} and \ref{TABLE3}, the improvements in 
compression results justify the computational overhead. 

\end{document}